\documentclass[apl,amsmath,amssymb,reprint,longbibliography,showkeys,superscriptaddress,balancelastpage]{revtex4-1}

\usepackage{textcomp}
\usepackage{graphicx}
\usepackage{dcolumn}
\usepackage{bm}
\usepackage{multirow}

\makeatletter
\def\NAT@def@citea{\def\@citea{\NAT@separator}}
\makeatother

\usepackage{color}
\usepackage{cleveref}
\crefformat{figure}{#2Fig.~#1#3}
\crefformat{equation}{#2Eq.~(#1)#3}
\crefformat{table}{#2Table~#1#3}
\crefformat{section}{#2Section~#1#3}

\begin{document}

\title{Thermal conductivity reduction in (Zr$_{0.25}$Ta$_{0.25}$Nb$_{0.25}$Ti$_{0.25}$)C high entropy carbide from extrinsic lattice defects}

\author{Cody A. Dennett}
	\email{cdennett@mit.edu}
	\affiliation{Materials Science and Engineering Department, Idaho National Laboratory, Idaho Falls, ID 83415, USA}
	\affiliation{Department of Nuclear Science and Engineering, Massachusetts Institute of Technology, Cambridge, MA 02139, USA}
\author{Zilong Hua}
	\affiliation{Materials Science and Engineering Department, Idaho National Laboratory, Idaho Falls, ID 83415, USA}
\author{Eric Lang}
	\affiliation{Center for Integrated Nanotechnologies, Sandia National Laboratories, Albuquerque, NM 87123, USA}
\author{Fei Wang}
	\affiliation{Department of Mechanical and Materials Engineering, University of Nebraska-Lincoln, Lincoln, NE 68588, USA}
\author{Bai Cui}
	\affiliation{Department of Mechanical and Materials Engineering, University of Nebraska-Lincoln, Lincoln, NE 68588, USA}
	\affiliation{Nebraska Center for Materials and Nanoscience, University of Nebraska-Lincoln, Lincoln, NE 68588, USA}

\date{\today}

\begin{abstract}
High entropy carbides ceramics with randomly-distributed multiple principal cations have shown high temperature stability, low thermal conductivity, and possible radiation tolerance. While chemical disorder has been shown to suppress thermal conductivity in these materials, little investigation has been made on the effects of additional, extrinsically-generated structural defects on thermal transport. Here, (Zr$_{0.25}$Ta$_{0.25}$Nb$_{0.25}$Ti$_{0.25}$)C is exposed to Zr ions to generate a micron-scale, structural-defect-bearing layer. The reduction in lattice thermal transport is measured using laser thermoreflectance. Conductivity changes from different implantation temperatures suggest dislocation loops contribute little to phonon scattering while nanoscale defects serve as effective scatterers, offering a pathway for thermal engineering.
\end{abstract}

\maketitle

\section{Introduction}

High entropy carbide ceramics (HECs) have emerged in the last several years as a promising class of materials for applications involving high temperatures or other environmental extremes. Borrowing from the maturing field of complexity-engineering and entropic-stabilization in metal alloys~\cite{Miracle2017}, new high entropy ceramics with complex selections of cations have emerged across oxides, borides, nitrides, sulfides, silicides, and carbides~\cite{Oses2020,Feng2021}. HECs, in particular, have been synthesized as bulk specimens and as thin films and have been shown to have a variety of physical characteristics attributed to the presence of a disordered cation sublattice including high hardness~\cite{Sarker2018,Ye2019,Harrington2019}, low thermal conductivity~\cite{Yan2018,Rost2020}, oxidation resistance~\cite{Zhou2018,Ye2019a,Ye2020}, and radiation tolerance~\cite{Wang2020}. 

To date, experimental studies of the thermal transport characteristics of these HEC materials have been relatively few and primarily focused on five-cation chemistries including (Hf$_{0.2}$Zr$_{0.2}$Ta$_{0.2}$Nb$_{0.2}$Ti$_{0.2}$)C~\cite{Yan2018,Wen2020,Rost2020} and (Hf$_{0.2}$Zr$_{0.2}$Ta$_{0.2}$Mo$_{0.2}$W$_{0.2}$)C~\cite{Rost2020}. These studies have exclusively focused on the thermal transport characteristics of as-synthesized materials in addition to their resulting structure. While some progress has been made in computationally treating transport in these systems, large discrepancies exist in the estimation of the lattice contribution to thermal conductivity between simulation and experiment~\cite{Dai2020}. While noted as a promising feature of these systems, the ability to thermally engineer HECs for targeted applications has not yet been exploited in a systematic manner.

In particular, the interplay between chemical disorder, stoichiometry, and \emph{structural defects} in relation to thermal transport has yet to be explored in HECs or, to a large extent, in any high entropy ceramic. Frameworks for understanding the effects of structural defects, and tailoring performance by extrinsically introducing such defects, are well-established for single-principal-element materials and compounds~\cite{Gurunathan2020,Khafizov2019}. Effects on electron and phonon thermal conductivity are routinely expressed in terms of the scattering strength of structural defects of particular types including 0D point defects, 1D dislocations, 2D dislocation loops or boundaries, and 3D inclusions~\cite{Alfred1966,Klemens1958,Turk1974,Morelli1993}. However, how thermal carriers scatter from each of these classes of structural features in the limit of maximized chemical disorder is not yet well understood. For HECs in particular, the relative partitioning of these thermal carriers (electrons and phonons) depends sensitively on stoichiometry~\cite{Rost2020} and introduces additional challenges for developing a complete understanding of transport phenomena. However, controlling carrier partitioning, therefore, could also eventually serve as a tool to engineer thermal properties. 

Here, the coupled effects of chemical disorder and extrinsically-imposed lattice defects are studied in (Zr$_{0.25}$Ta$_{0.25}$Nb$_{0.25}$Ti$_{0.25}$)C. This four-cation HEC was originally synthesized Hf-free by Wang and coworkers to study its potential, if any, for use in nuclear systems, as Hf has a large neutron absorption cross section~\cite{Wang2020}. Specimens of (Zr$_{0.25}$Ta$_{0.25}$Nb$_{0.25}$Ti$_{0.25}$)C were exposed to Zr ion beam irradiation to generate a micron-scale defect-bearing layer at the surface of bulk specimens at a series of different temperatures. Following exposure, a spatial domain thermal reflectance (SDTR) method is used to determine the total thermal conductivity of the defect-bearing region using a multi-layer thermal transport model. A combination of SDTR and electrical resistivity measurements on a pristine bulk HEC are used to determine the total heat capacity and estimate the electron contribution to thermal conductivity. The resulting thermal performance is interpreted in the context of previously-measured concentrations of defects from electron microscopy, showing that dislocation loops serve as much less effective thermal scatters in these chemically disordered structures than would be expected from classical phonon scattering theory.

\section{Materials and Methods}

The (Zr$_{0.25}$Ta$_{0.25}$Nb$_{0.25}$Ti$_{0.25}$)C specimens investigated in this work have been described previously \cite{Wang2020}, so only a brief summary of their synthesis and ion beam processing will be described here. Bulk specimens were fabricated from commercial powders of the constituent binary carbides through ball milling the as-received powders and consolidation with spark plasma sintering (SPS). Scanning electron microscopy indicated an as-synthesized mean grain size of 19~{\textmu}m. X-ray diffraction patterns of as-synthesized specimens are indexed as a single-phase rock-salt structure, similar to other multi-cation carbides~\cite{Wang2020}. The theoretical density is calculated from the XRD lattice constant as 8.46~g/cm$^3$ and the measured density of the bulk materials is 8.25~g/cm$^3$, resulting in a relative density of 97.5\%. Mechanically polished specimens were exposed to 3~MeV Zr$^{2+}$ ions to a total fluence of 8.0$\times$10$^{15}$~ions/cm$^2$ at three temperatures, 25, 300, and 500$^\circ$C. Ions at this energy generate structural defects over an $\sim$1.5~{\textmu}m depth at the sample surface, with peak defect production at $\sim$800~nm~\cite{Ziegler2010,Weber2019}. Extensive transmission electron microscopy (TEM) characterization by Wang et al. showed the generation of a population of both perfect and faulted Frank loops with consistently small diameters ($\sim$2~nm) at all three temperatures, with density peaking near the peak defect generation region, and no other visible defects~\cite{Wang2020}. A spatially-averaged summary of the measured dislocation loop size and density is listed in \cref{tab:loops}, noting that no loops were observed in the as-synthesized material. No radiation-induced chemical segregation was observed near the grain boundaries for the as-synthesized or irradiated materials. 

\begin{table}	
\centering
\begin{tabular}{cccc}\hline
\begin{tabular}{l}Irradiation \\ Temperature\end{tabular} & \begin{tabular}{l}Ion \\ Fluence\end{tabular} & \begin{tabular}{l}Ave. Loop \\ Diameter [$d_l$]\end{tabular} & \begin{tabular}{l}Ave. Loop \\ Density [$n$]\end{tabular} \\ \hline
25$^\circ$C & 8$\times$10$^{15}$~ions/cm$^2$ & 1.6~nm & 9.5$\times$10$^{17}$~1/cm$^3$ \\
300$^\circ$C & 8$\times$10$^{15}$~ions/cm$^2$ & 2.0~nm & 7.4$\times$10$^{17}$~1/cm$^3$ \\
500$^\circ$C & 8$\times$10$^{15}$~ions/cm$^2$ & 2.1~nm & 6.0$\times$10$^{17}$~1/cm$^3$ \\ \hline
\end{tabular}
\caption{Measured dislocation loop diameter and density in Zr-ion irradiated HECs at different temperatures from Wang and coworkers~\cite{Wang2020}. Averages of the spatially-heterogeneous loop microstructure are given here as a relative comparison between different irradiation temperatures.\label{tab:loops}}
\end{table}

For both pristine and ion irradiated (Zr$_{0.25}$Ta$_{0.25}$Nb$_{0.25}$Ti$_{0.25}$)C, SDTR is used to extract thermal properties including thermal conductivity, $\kappa$, thermal diffusivity, $D$, and, in the case of the as-synthesized material, the specific heat capacity, $C_p$. In SDTR, periodic local temperature variations are induced using an intensity-modulated 660~nm CW heating laser and detected using a 532~nm CW probing laser through the thermoreflectance effect~\cite{Hurley2015,Khafizov2017}. The heating and probing lasers are both focused to spot sizes on the order $\sim$1~{\textmu}m using a high-power optical objective and the probing laser scanned across a distance of $\sim$20~{\textmu}m centered on the heating spot. By coating specimens with thin transducer layers of gold, a strong thermoreflectance response at 532~nm is ensured. The relative phase lag between the heating and probe laser as a function of separation distance at several modulation frequencies in the range 5--100~kHz is recorded using lock-in detection~\cite{Hurley2015}. A multi-layer thermal model taking into account the thermal properties of an arbitrary number of layers using the thermal quadrupole method is used to return best-fit values for the properties of selected layers~\cite{Maillet2000,Hua2019}.

\begin{figure}[b]
\centering
\includegraphics[width=0.45\textwidth]{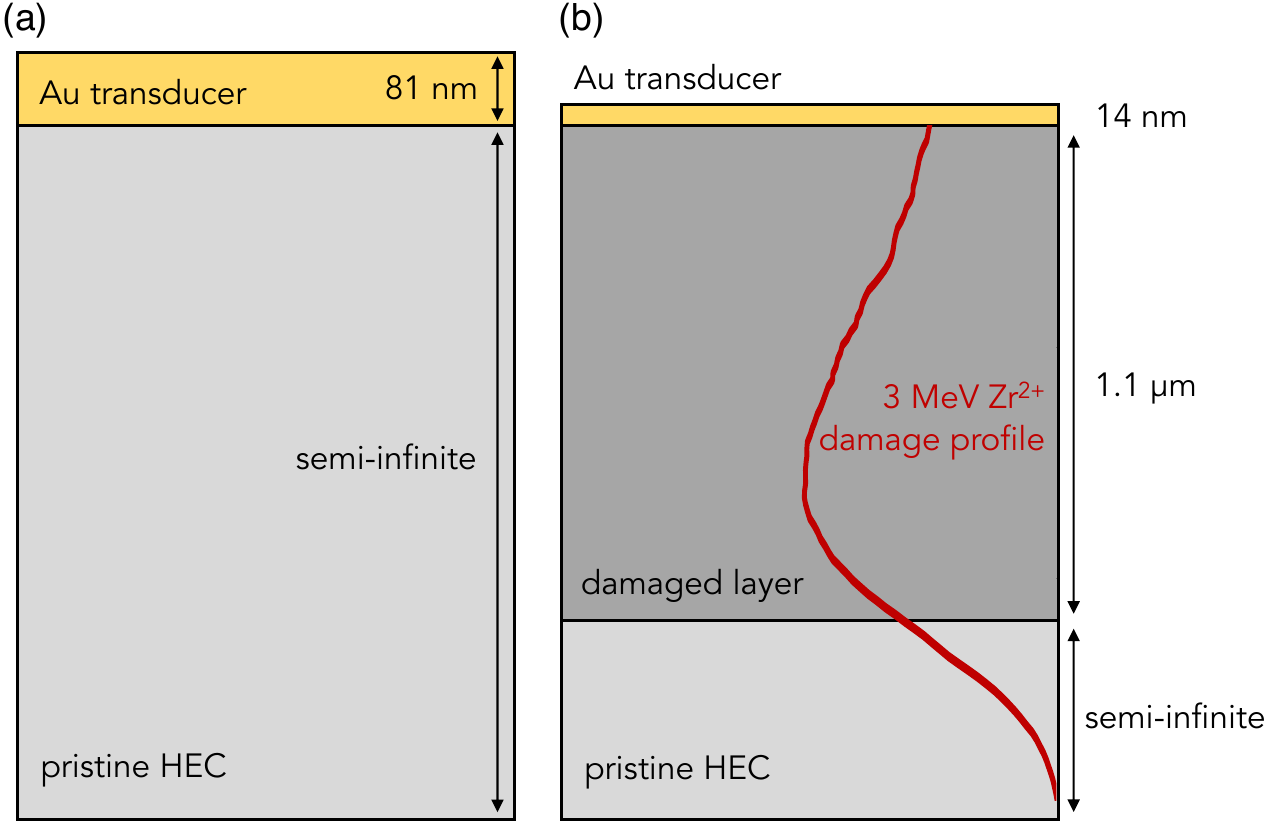}
\caption{Thermal layering model used for pristine (a) and ion irradiated (b) HECs, overlaid with the defect generation profile from Zr ion implantation. The thick gold transducer layer used on the pristine specimen allows for all three thermal parameters, $\kappa$, $D$, and $C_p$, to be extracted from the two-layer thermal model. The defect generation profile and ion irradiated layer are to scale, while the gold transducer thicknesses are not.\label{fig:layers}}
\end{figure} 

The thermal layering model used for pristine and as-irradiated (Zr$_{0.25}$Ta$_{0.25}$Nb$_{0.25}$Ti$_{0.25}$)C is shown in \cref{fig:layers}. For pristine specimens, \cref{fig:layers}(a), a relatively thick 81~nm Au film was deposited in order to achieve the highest joint sensitivity to  thermal conductivity and thermal diffusivity based on analysis shown in the Supplementary Material. With known density, independent optimization of both $\kappa$ and $D$ allows $C_p$ to be determined as $C_p=\kappa/D\rho$, where $\rho$ is the mass density. For all SDTR analysis, the theoretical density is fixed for analysis as scans on the order of 20~{\textmu}m remain largely intragranular. In addition, differential scanning calorimetry (DSC) measurements of $C_p$ are made independently on pristine (Zr$_{0.25}$Ta$_{0.25}$Nb$_{0.25}$Ti$_{0.25}$)C to confirm the value inferred through SDTR.

For ion-irradiated samples, \cref{fig:layers}(b) shows the three-layer thermal model overlaid with spatially-varying SRIM-calculated profile of displacement damage. Previous investigations of ion-implanted materials using SDTR and other thermoreflectance approaches have proposed various methods for the determining the most appropriate layer segmentation for continuously spatially heterogeneous material volumes~\cite{Hua2019,Riyad2018}. Over-segmenting this defect-bearing layer leads to an ill-posed inverse problem, causing issues of uniqueness and sensitivity in the fitted solution. As such, a three-layer thermal model was used here with the thickness of the ion-modified layer fixed at 1.1~{\textmu}m for all conditions. This distance corresponds to half the maximum value of the displacement damage near the end of range. For the three-layer model, $C_p$ is assumed constant for both defect-bearing and pristine layers, as there is little evidence to suggest that heat capacity is meaningfully affected by structural defects. In each optimization, the thermal parameters of the pristine substrate layer are fixed to those measured independently on the pristine specimen and $D$ and $\kappa$ optimized self-consistently for the defect-bearing layer. A table of the fixed and varying optimization parameters used in the two- and three-layer models are detailed in the Supplementary Materials. 

\begin{figure}
\centering
\includegraphics[width=0.45\textwidth]{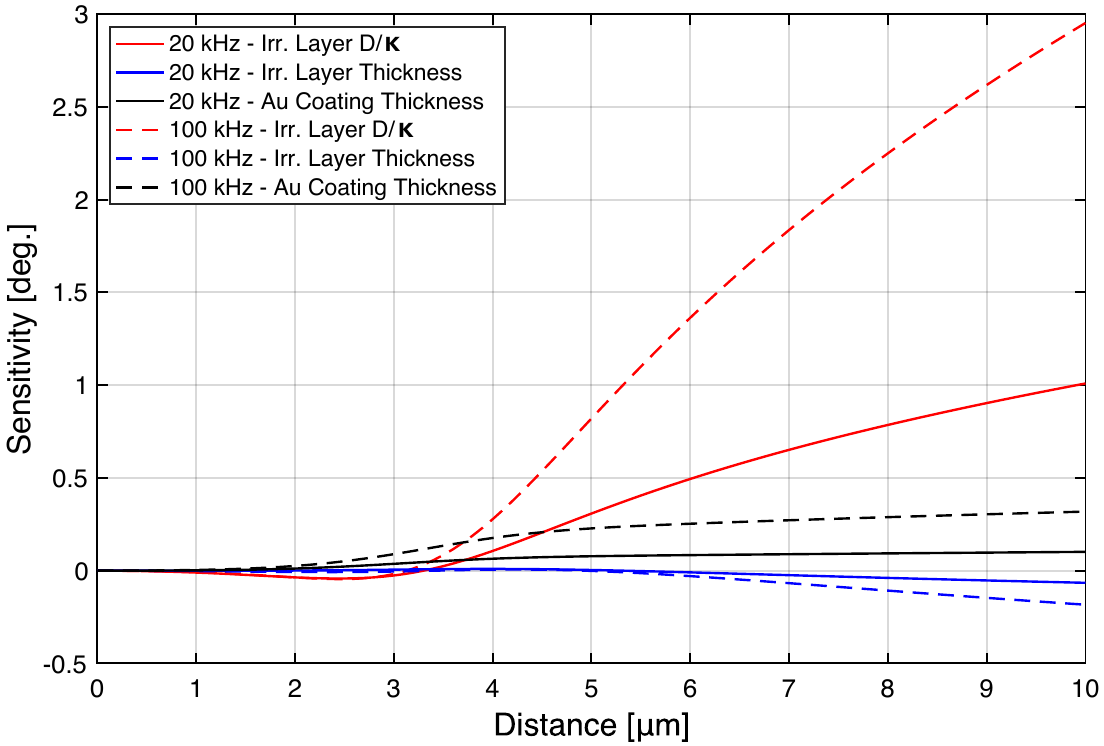}
\caption{Sensitivity of the three-layer thermal model to several parameters at both 20~kHz and 100~kHz thermal wave modulation frequencies in terms of phase delay difference as a function of SDTR scan distance, the distance between heating and probe laser spots. Initial parameters are taken as 14~nm of Au coating, fixed values of semi-infinite substrate thermal properties, and a 10\% reduction in conductivity in the middle, defect-bearing layer. To ensure a constant heat capacity, the values of $D$ and $\kappa$ in the defected layer are changed in the same proportion in this analysis. Of these parameters, the highest sensitivity is to the thermal properties of the layer of interest.\label{fig:sens}}
\end{figure}

To ensure this parameterization results in the highest sensitivity to the thermal properties of the defect-bearing layer, a detailed sensitivity analysis is carried out, see \cref{fig:sens}. Relative sensitivity is calculated for $D$ and $\kappa$ of the irradiated layer, the thickness of the irradiated layer, and the thickness of the gold transducer film as 
\begin{equation}
S(\xi)=\frac{\varphi(\xi+\Delta\xi)-\varphi(\xi)}{(\Delta\xi / \xi)},
\end{equation}
where $\varphi(\xi)$ is the spatially-varying thermal phase delay and $\xi$ is the parameter in question. \cref{fig:sens} is calculated for a baseline set of parameters $D=3.70$~mm$^2$/s and $\kappa=9.14$~W/m$\cdot$K for the pristine materials, with a 10\% reduction in $D$ and $\kappa$ in the defect-bearing layer, and assuming a 14~nm thick deposited gold film (with corresponding thermal conductivity~\cite{Dennett2020,Chen1999}). An $\sim$10~nm gold film was chosen as the set point as the sensitivity to a ``sandwiched'' thermal layer is higher when the transducer layer is thin; 14~nm was eventually deposited on these irradiated samples. As can be seen in \cref{fig:sens}, the highest sensitivity at both the intermediate and highest modulation frequencies used, 20~kHz and 100~kHz, is to the thermal properties of the defect-bearing layer. Only very weak sensitivities to the precise gold coating thickness and segmented irradiated layer thickness are observed. Together, these sensitivities ensure that the values reported from the three-layer optimization are representative of the defect-modified thermal characteristics.  

Total thermal conductivity as measured in these carbide ceramics contains contributions from both electron and lattice (or phonon) thermal transport~\cite{Rost2020}. To extract the relative contribution of electrons to thermal conductivity, the electrical resistivity of the un-coated side of the pristine bulk (Zr$_{0.25}$Ta$_{0.25}$Nb$_{0.25}$Ti$_{0.25}$)C specimen was measured using a 4-point probe. Traditionally, the electrical resistivity can be used to estimate the electronic thermal conductivity as $\kappa_e=LT/\rho_e$, where most commonly $L=L_0=2.44\times10^{-8}$~W$\Omega$K$^{-2}$ is the Lorentz number and $\rho_e$ is the electrical resistivity. However, in single-element metal carbides, $L_0$ has been shown to overestimate $\kappa_e$, such that an effective Lorentz number as calculated by Makinson should be used~\cite{Rost2020,Makinson1938,Zheng2017}. A description of the 4-point probe measurements as well as the conversion of $\rho_e$ to $\kappa_e$ is provided in the Supplementary Materials. From $\kappa_e$ and the measured total thermal conductivity, $\kappa_{\mathrm{tot}}$, the phonon thermal conductivity can be estimated as $\kappa_p=\kappa_{\mathrm{tot}}-\kappa_e$. The relative electron contribution to $\kappa_{\mathrm{tot}}$ has been shown to be a sensitive function of carbon stoichiometry, and therefore bonding character, in binary carbides and these HECs~\cite{Rost2020,Hossain2021}. The carbon stoichiometry of as-synthesized (Zr$_{0.25}$Ta$_{0.25}$Nb$_{0.25}$Ti$_{0.25}$)C is estimated using X-ray photoemission spectroscopy (XPS) to place measured values of $\kappa_e$ in context. As extraction of the electrical resistivity of solely the defect-bearing layers in the irradiated materials was not possible from bulk specimens, $\kappa_e$ is assumed to be constant for the defect-bearing material for this initial investigation. This practical limitation will lead to an overestimation of the lattice thermal conductivity reduction, as both electrons and phonons are known to scatter from point defects in refractory carbide and nitride ceramics~\cite{Zheng2017}. 

\section{Results and Discussion}

SDTR measurements of pristine (Zr$_{0.25}$Ta$_{0.25}$Nb$_{0.25}$Ti$_{0.25}$)C with a thick gold transducer return room-temperature thermal property values of $k_{\mathrm{tot}}=9.14\pm0.20$~W/m$\cdot$K and $D=3.70\pm0.06$~mm$^2$/s, which imply a specific heat capacity of $C_p=293\pm4$~J/kg$\cdot$K, with all uncertainties listed as the standard error of multiple spatially-varying measurements. Although no direct comparisons exist for this 4-cation HEC in the literature, these values are consistent with the thermal properties measured in bulk (Hf$_{0.2}$Zr$_{0.2}$Ta$_{0.2}$Nb$_{0.2}$Ti$_{0.2}$)C synthesized using the same procedure~\cite{Yan2018}. \cref{tab:thermal_prop} compares the as-measured values of the 4- and 5-cation HEC thermal properties. That the 4-cation HEC measured here retains a higher thermal conductivity is consistent with the lower levels of mass and force constant scattering expected with one fewer principal elements. In addition, the distinctly higher value for $C_p$ is consistent with excluded cation, Hf, having the lowest $C_p$ of any of the constituent binary carbides~\cite{Yan2018}. Confirmation measurements of $C_p$ made using DSC returned a value $310\pm5$~J/kg$\cdot$K at $\sim$335~K, consistent with the theoretically monotonically-rising heat capacity in refractory carbides in this temperature range~\cite{Wolf1999}. 

\begin{table}
\begin{centering}
\begin{tabular}{cccc}\hline
 & $D$ [mm$^2$/s] & $\kappa_{\mathrm{tot}}$ [W/m$\cdot$K] & $C_p$ [J/kg$\cdot$K] \\ \hline
4-cation$^1$ & 3.70 & 9.14 & 293 \\
5-cation$^2$ & 3.60 & 6.45 & 191 \\ \hline
\end{tabular}\\
$^1$(Zr$_{0.25}$Ta$_{0.25}$Nb$_{0.25}$Ti$_{0.25}$)C\\
$^2$(Hf$_{0.2}$Zr$_{0.2}$Ta$_{0.2}$Nb$_{0.2}$Ti$_{0.2}$)C~\cite{Yan2018}

\end{centering}
\caption{Comparison of thermal properties for 4- and 5- component bulk HECs. $D$ and $\kappa$ are measured for the 4-cation HEC using SDTR and the resulting $C_p$ calculated, while Yan and coworkers used laser flash to measure $D$, DSC to measure $C_p$, calculated $\kappa$ for the 5-cation HEC.\label{tab:thermal_prop}}
\end{table}

4-point probe measurements indicated that pristine (Zr$_{0.25}$Ta$_{0.25}$Nb$_{0.25}$Ti$_{0.25}$)C specimens have an electrical resistivity $\rho_e=149\pm2$~{\textmu}$\Omega$$\cdot$cm. This is higher than that of binary carbides such as TiC and TaC (68 and 25~{\textmu}$\Omega$$\cdot$cm, respectively~\cite{Williams1999}), similar to that for (Hf$_{0.2}$Zr$_{0.2}$Ta$_{0.2}$Nb$_{0.2}$Ti$_{0.2}$)C reported as $\sim$115~{\textmu}$\Omega$$\cdot$cm by Wen and coworkers for sintered pellets~\cite{Wen2020}, but lower than the range reported for thick films of (Hf$_{0.2}$Zr$_{0.2}$Ta$_{0.2}$Mo$_{0.2}$W$_{0.2}$)C$_{1-x}$ by Rost and coworkers as a function of carbon stoichiometry as $\sim$200--1200~{\textmu}$\Omega$$\cdot$cm~\cite{Rost2020}. That the measured resistivity is low compared to the range reported for precisely carbon-controlled HECs implies the final, as-synthesized bulk samples are possibly carbon deficient and retain some metallic bonding that would not be present in a perfectly stoichiometric compound. From $\rho_e$, $\kappa_e$ is estimated as 3.7~W/m$\cdot$K, indicating that 40\% of the thermal conductivity is due to electrons in the as-synthesized material and 60\% is due to the lattice. From XPS measurements, the metal to carbon ratio is M:C$\simeq$1, as detailed in the Supplementary Materials, indicating that an approximately stoichiometric carbide has been maintained. While Rost and coworkers reported generally higher resistivity for stoichiometric compounds, the deposition of excess, non-metal-bonded carbon in sputter deposited materials complicates a direct comparison~\cite{Rost2020}. Further correlated microscopy and XPS should be carried out in the future on bulk, sintered ceramics to more fully characterize possibly heterogeneous carbon stoichiometry following synthesis. To ensure that thermal property measurements account for any such variability, ten or more randomly-selected spatially varying measurements are made on each specimen prior to reporting statistical averages.

\begin{figure}
\centering
\includegraphics[width=0.45\textwidth]{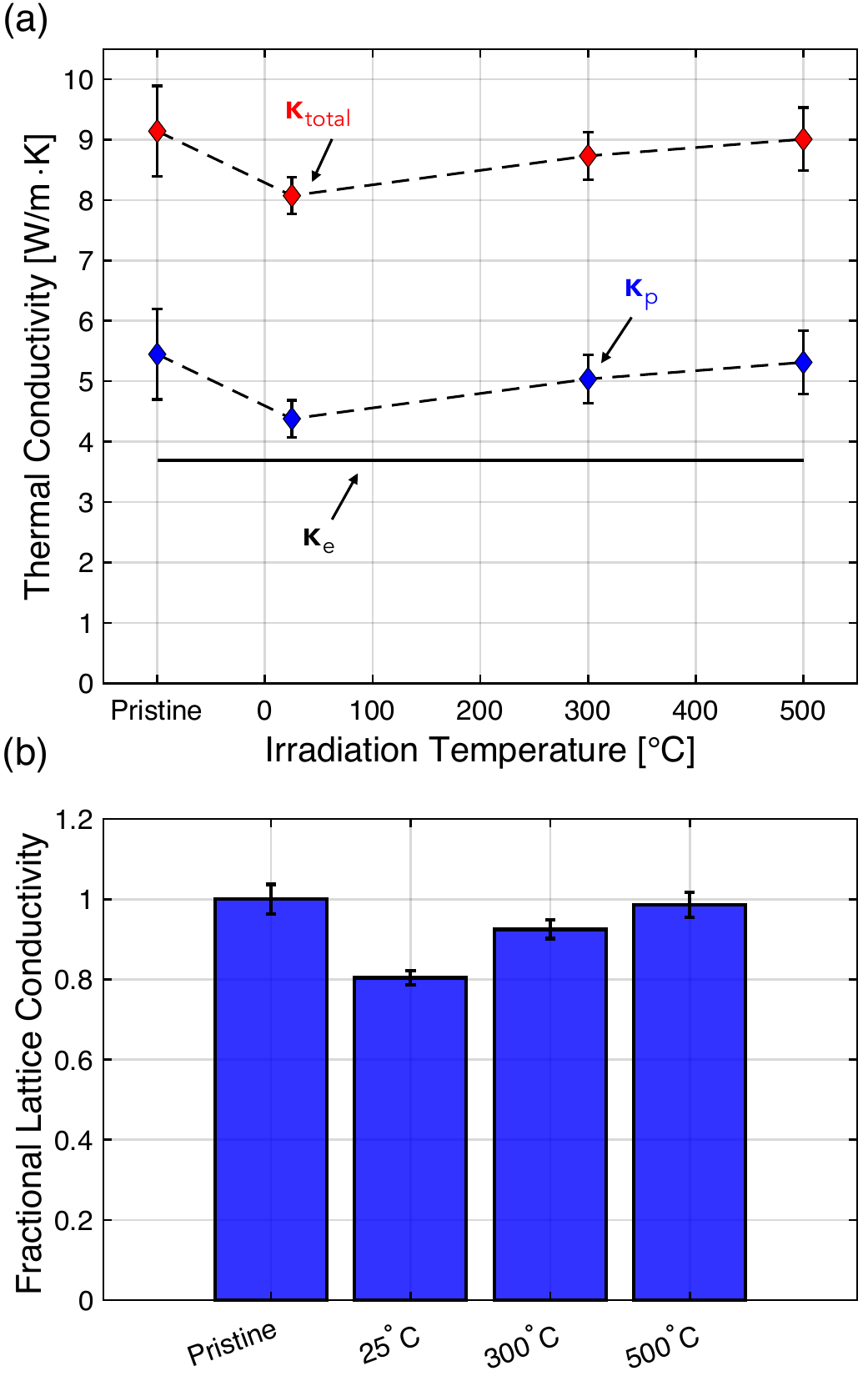}
\caption{(a) Total, electronic, and phonon thermal conductivity for the pristine HEC and at each of the three ion exposure temperatures. A constant electronic thermal conductivity, $\kappa_e$, as measured on the pristine specimen is subtracted from the measured $\kappa_{\textrm{total}}$ to recover $\kappa_p$. (b) Fractional lattice conductivity retained at each ion irradiation temperature. As highlighted, approximately 20\% of the lattice thermal conductivity is lost when irradiated at 25$^\circ$C to this fluence level, while the majority of the lattice conductivity is retained for 500$^\circ$C irradiations. Error bars in (a) are shown as the standard deviation, $\sigma$, of $N=10$ or more spatially-varying measurements per condition to show the spread in collected data and in (b) are shown as the standard error, $\sigma/\sqrt{N}$, of that sampling set to reflect the overall confidence in the measured value. \label{fig:cond}}
\end{figure}

Using the three-layer thermal transport model as described above, \cref{fig:cond}(a) shows the total and phonon thermal conductivities as a function of irradiation temperature for the fixed ion fluence applied to each specimen. The electronic contribution is plotted as the constant value estimated for the as-synthesized material. The general trend of a larger decrease in thermal conductivity at lower irradiation temperatures is consistent with expectations from most material systems where extrinsic structural defect production is driven by radiation bombardment. Namely, higher irradiation temperatures increase the mobility and therefore the recombination of any Frenkel defects or small clusters generated by displacement damage, leaving fewer thermal scattering sites~\cite{Khafizov2014,Ferry2019,Dennett2020}. Using the simplified assumption that $\kappa_e$ is unaffected by structural defects and the entire reduction is due to phonon scattering, the fractional reduction in lattice thermal conductivity is plotted in \cref{fig:cond}(b). With this simplified assumption that $\kappa_e$ is unaffected by radiation-generated defects, an $\sim$20\% reduction in $\kappa_p$ is observed at 25$^\circ$C, with very little, if any, reduction for exposure at 500$^\circ$C. 

The behavior of the lattice thermal conductivity in the presence of these structural defects is most easily explored using the simplified Klemens-Callaway-Debye model where
\begin{equation}\label{eq:KC}
\kappa_p=\frac{1}{3}\int_0^{\omega_{\mathrm{max}}}d\omega C(\omega)v_g(\omega)^2\tau(\omega),
\end{equation}
and $C(\omega)$, $v_g(\omega)$, and $\tau(\omega)$ are the phonon-frequency-dependent heat capacity, group velocity, and relaxation time, respectively~\cite{Toberer2011}. The overall relaxation time is normally considered as a combination of different scattering mechanisms, phonon-phonon scattering, impurity scattering, etc., as summed using Matthiessen's rule. Classical models have been used with success to describe scattering from both perfect (pl) and faulted (fl) dislocation loops in terms of the density and average size of the dislocation loops present as
\begin{equation}
\tau_{pl}^{-1}\propto n d_l\omega^2,
\end{equation}
and
\begin{equation}
\tau_{fl}^{-1}\propto n d_l^2\omega^2,
\end{equation}
where $n$ is the dislocation loop density and $d_l$ is the loop diameter, with each relationship containing prefactors, not shown here, to maintain the correct dimensionality~\cite{Khafizov2019}. As Wang and coworkers showed that some number of both perfect and faulted loops were present in these specimens~\cite{Wang2020}, it is useful to compare the products $nd_l$ and $nd_l^2$ to compare the relative effects of extrinsic dislocation loops on $\kappa_p$. This comparison is shown in \cref{tab:scattering} and demonstrates that scattering from either perfect or faulted dislocation loops should be nearly identical in each of the three specimens with extrinsic structural defects measured here.

\begin{table}
\centering
\begin{tabular}{ccc}\hline
\begin{tabular}{l}Irradiation \\ Temperature\end{tabular} & Perfect [$nd_l$] & Faulted [$nd_l^2$] \\ \hline
25$^\circ$C & 1.52$\times$10$^{-3}$~1/nm$^2$ & 2.43$\times$10$^{-3}$~1/nm \\
300$^\circ$C & 1.48$\times$10$^{-3}$~1/nm$^2$ & 2.96$\times$10$^{-3}$~1/nm \\
500$^\circ$C & 1.26$\times$10$^{-3}$~1/nm$^2$ & 2.64$\times$10$^{-3}$~1/nm \\ \hline
\end{tabular}
\caption{Comparison of the phonon scattering rate dependency on dislocation loop size and density for both perfect and faulted dislocation loops. Both scattering rates share a common $\omega^2$ phonon frequency dependence.\label{tab:scattering}}
\end{table}

This theoretically identically scattering strength from loops, coupled with the similarly in thermal conductivity between the pristine sample and that exposed at 500$^\circ$C, implies that dislocation loops are effectively not scattering phonons in this chemically complex ceramic. This conclusion is reached by noting that at 500$^\circ$C, few residual point defects are expected following irradiation given the increased defect mobility at high temperatures. These data suggest, therefore, that the interactions between phonons and dislocation loops is inherently distinct in materials with a high degree of chemical complexity compared with traditional ceramics.
Analytical dislocation loop scattering models, after Klemens, propose that the long-range strain field associated with dislocations is responsible for the majority of phonon scattering~\cite{Klemens1955,Klemens1958}. Here, in a maximally-chemically disordered carbide ceramic, dislocations appear to contribute little to phonon scattering, implying their strain fields are suppressed.t This long-range suppression could occur as a result of the inherit strain associated with the disordered cation sublattice, although further experimental and computational work is required to investigate these effects directly. 

Even though the reduction in $\kappa_p$ is likely overestimated here due to assumption of constant $\kappa_e$ across exposure conditions, electrons scatter little from extended loops. Therefore, the conclusion that residual nanoscale structural defects are the primary contributors to conductivity reduction remains unchanged. Altogether, that lattice thermal conductivity is reduced by $\sim$20\% at 25$^\circ$C, in a microstructure expected to contain a high density of both nanoscale and extended defects, supports the conjecture that heat carrier scattering from extrinsic nanoscale defects (point defects and small clusters below TEM resolution) can be used as a tool with which thermal conductivity may be engineered in these systems. Although ion beam irradiation was used to generate extrinsic structural defects in this work, other post-synthesis processing treatments could similarly be used for such defect injection and thermal engineering in the future.

\section{Conclusion}

Here, the impact of lattice defects extrinsically-generated using ion irradiation on thermal transport in a chemically complex carbide ceramic with four principal cations has been systematically studied using a laser thermoreflectance technique. Multi-layer thermal modeling allows the defect-affected thermal conductivity of a micron-thick damaged layer generated by ion implantation to be measured. Additional electrical resistivity measurements on as-synthesized material shows that this HEC has a thermal carrier partitioning, between electrons and phonons, in line with that previously observed in similar materials. The reduction in thermal conductivity is observed to be greatest in materials irradiated at low temperatures where a significant population of retained nanoscale defects is expected. At high temperatures, however, little reduction in overall or phonon thermal conductivity is observed. Coupled with prior investigations showing that small dislocation loops have formed under these exposure conditions at all temperatures, these observations indicate that dislocation loops are contributing little to phonon scattering, likely due to a suppression of their long-range strain fields. Nanoscale defects still serve as effective scatterers of thermal carriers in this system, opening a pathway for controlled thermal transport engineering in the presence of maximized chemical disorder in high entropy ceramics.

\section{Acknowledgments}
The authors would like to thank S.G. Rosenberg and M. Meyerman at SNL for performing XPS measurements. This work was supported through the INL Laboratory Directed Research \& Development Program under U.S. Department of Energy Idaho Operations Office Contract DE-AC07-05ID14517. C.A.D. and Z.H. acknowledge support from the Center for Thermal Energy Transport under Irradiation (TETI), an Energy Frontier Research Center funded by the US Department of Energy, Office of Science, Office of Basic Energy Sciences. This work was performed, in part, at the Center for Integrated Nanotechnologies, an Office of Science User Facility operated for the DOE Office of Science. Sandia National Laboratories is a multimission laboratory managed and operated by National Technology \& Engineering Solutions of Sandia, LLC, a wholly owned subsidiary of Honeywell International, Inc., for the U.S. DOE's National Nuclear Security Administration under contact DE-NA-0003525. The views expressed in this article do not necessarily represent the views of the U.S. DOE of the United States Government.


\bibliography{ref}

\end{document}


\title[]{Supplementary Materials:\\ ~\vspace{-0.75pc} \\Thermal conductivity reduction in (Zr$_{0.25}$Ta$_{0.25}$Nb$_{0.25}$Ti$_{0.25}$)C high entropy carbide from extrinsic lattice defects}

\author{Cody A. Dennett}
	\email{cdennett@mit.edu}
	\affiliation{Materials Science and Engineering Department, Idaho National Laboratory, Idaho Falls, ID 83415, USA}
	\affiliation{Department of Nuclear Science and Engineering, Massachusetts Institute of Technology, Cambridge, MA 02139, USA}
\author{Zilong Hua}
	\affiliation{Materials Science and Engineering Department, Idaho National Laboratory, Idaho Falls, ID 83415, USA}
\author{Eric Lang}
	\affiliation{Center for Integrated Nanotechnologies, Sandia National Laboratories, Albuquerque, NM 87123, USA}
\author{Fei Wang}
	\affiliation{Department of Mechanical and Materials Engineering, University of Nebraska-Lincoln, Lincoln, NE 68588, USA}
\author{Bai Cui}
	\affiliation{Department of Mechanical and Materials Engineering, University of Nebraska-Lincoln, Lincoln, NE 68588, USA}
	\affiliation{Nebraska Center for Materials and Nanoscience, University of Nebraska-Lincoln, Lincoln, NE 68588, USA}

\maketitle

\section*{Gold Film Thickness for Pristine HEC}

To determine the thickness of the Au coating to use for the as-synthesized (Zr$_{0.25}$Ta$_{0.25}$Nb$_{0.25}$Ti$_{0.25}$)C to provide the highest degree of sensitivity to $C_p$, a detailed sensitivity analysis is carried out using the previously-measured values for the thermal parameters of (Hf$_{0.2}$Zr$_{0.2}$Ta$_{0.2}$Nb$_{0.2}$Ti$_{0.2}$)C for the two-layer thermal model as shown in Fig.~1(a) in the main manuscript, as the thermal parameters of the 4-cation HEC under investigation here had not been measured previously. As listed in the main body of the manuscript, these test values are: $\kappa=6.45$~W/m$\cdot$K and $D=3.60$~mm$^2$/s~\cite{Yan2018}. For values of Au coating thickness between 10 and 200~nm, the sensitivity defined as $S(\xi)=[\varphi(\xi+\Delta\xi)-\varphi(\xi)]/[\Delta\xi / \xi]$, where $\varphi$ is the thermal phase lag and $\xi$ is the parameter of interest ($\kappa_s$ or $D_s$), at a fixed distance of 10~{\textmu}m from the heating laser spot was calculated for modulation frequencies of 20 and 100~kHz (the middle and highest modulation frequencies used). In general, the sensitivity function need not be maximized at the furthest scan distance for all parameters in the thermal model~\cite{Hua2019}, but for Au coating thickness, this sensitivity maximum is always located at the furthest scan distance. For all cases, the convolved laser heating spot is taken at 1.8~{\textmu}m and the interfacial thermal resistance between the Au coating and the HEC substrate is taken as $R^{th}=4\times10^{-8}$~m$^2$K/W, consistent with values regularly measured in experiment. As the thermal conductivity of thin gold films is known to vary as a function of film thickness, $\kappa_f$ and $D_f$ for Au films at each thickness are computed using the model of Chen and coworkers~\cite{Chen1999}, fit to a series of films of different thicknesses all coated using the same DC magnetron sputter coater used in this study as described in the Supplementary Material of Ref.~\cite{Dennett2020}.

\begin{figure}[h]
\centering
\includegraphics[width=0.55\textwidth]{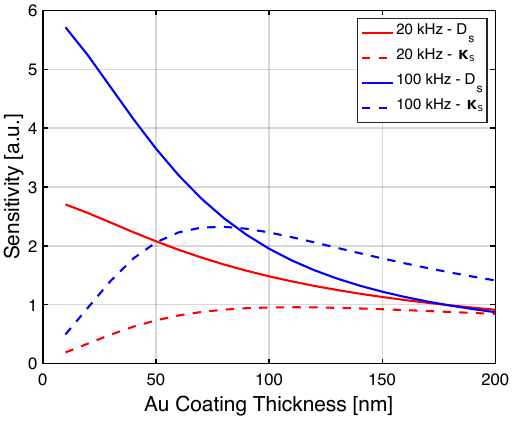}
\caption{Sensitivity of two-layer thermal model to Au coating thickness seeded with pristine HEC thermal properties. As sensitivity to $D_s$ decreases monotonically with Au coating thickness, the highest sensitivity to $C_p$ of the substrate should be near where the sensitivity to $\kappa_s$ peaks.\label{fig:p_sens}}
\end{figure}

The results of this sensitivity analysis re-computed with the final values optimized for (Zr$_{0.25}$Ta$_{0.25}$Nb$_{0.25}$Ti$_{0.25}$)C, $\kappa_s=9.14$~W/m$\cdot$K and $D_s=3.70$~mm$^2$/s, are shown in \cref{fig:p_sens}. The general features match those computed for the values of the 5-component HEC used originally and confirm that appropriate choices were made for the 4-cation HEC under investigation. For both 20 and 100~kHz modulation frequencies, the overall sensitivity to the thermal diffusivity of the substrate layer decreased monotonically as the thickness of the gold transducer layer increases. For the thermal conductivity, a broad peak in sensitivity at around 100~nm of Au is observed at both modulation frequencies. As the eventual determination of $C_p=\kappa/D\rho$ for the as-synthesized HEC requires both $\kappa_s$ and $D_s$ to be optimized individually, the sensitivity of both thermal parameters must be considered when choosing the thickness of the Au coating to apply. As the sensitivity to $\kappa_s$ increases rapidly until $\sim$80~nm, this thickness was chosen as the target for the as-synthesized HEC specimen, where the final deposited thickness is measured as $81\pm1$~nm using an optical transmission method~\cite{Hurley2015}. While \cref{fig:p_sens} shows this sensitivity only for the 4-cation HEC, the sensitivities calculated for the 5-cation HEC follow very similar trends and were used in the determination of the target thickness for Au coating.

\section*{SDTR Thermal Wave Data}

The multi-layer thermal transport model used to model the SDTR response, and therefore extract thermal transport properties, contains complex dependencies on the thermal parameters (thermal diffusivity -- $D$ and thermal conductivity -- $\kappa$) of each modeled layer as well as the interfacial thermal resistance -- $R^{th}$ -- between each layer and the convolved spot size of the pump and probe lasers -- $R_s$~\cite{Hurley2015}. The inverse problem of optimizing thermal parameters to the observed thermal phase lag becomes ill-posed if too many of these variable are allowed to vary in any given optimization. As such, the parameter spaces have been narrowed through independent property measurements for each of the two cases considered, a thick gold film on a pristine specimen and a thin gold film on a defect-bearing specimen, in order to retain the highest sensitivity to the parameters of interest. The set of fixed/varying parameters for the two- and three-layer cases used here are shown in \cref{tab:layers}. In both cases, $R_s$ and $R^{th}_1$ are allowed to vary as the laser focus and pre-coating surface conditions which lead to these values may vary point-to-point. For the 2-layer case, additional independent optimization of $D_2$ and $\kappa_2$ in the pristine specimen allow the heat capacity of the as-synthesized material to be determined as the theoretical density is known. For the 3-layer case, $D_2$ and $\kappa_2$ are optimized self-consistently, fixing $\rho C_p$ to that measured on the pristine specimen. In addition, the pristine layer beneath the ion penetration depth is fixed with the thermal properties of the as-synthesized material and the thermal resistance between the defect-bearing and pristine substrate, $R^{th}_2$, is set to 0 as this interface is, in reality, diffuse. In both cases, the thickness of the ``bulk'', or semi-infinite, substrate is set as 1~mm, orders of magnitude longer than the relevant thermal wave penetration depth. Parametric studies showed that on the scale of millimeters, the precise choice of substrate thickness in the model has no impact on the optimized result.

\begin{figure}
\centering
\includegraphics[width=\textwidth]{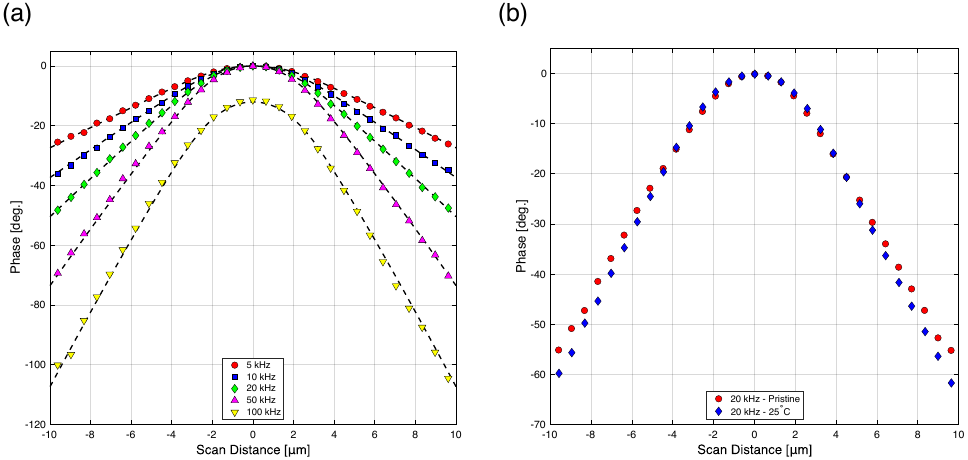}
\caption{(a) Experimental SPTR thermal wave profile for the pristine HEC shown at all five measurement frequencies. The offset of the highest-frequency, 100~kHz, data is retained in order to extract both thermal conductivity and diffusivity from a single measurement after Hurley and coworkers~\cite{Hurley2015}. (b) Comparison of recorded thermal wave phase profiles for pristine and 25$^\circ$C irradiated HECs at 20~kHz. The clear phase difference in the thermal far field indicates a reduction in thermal transport in the irradiated layer.\label{fig:TCM}}
\end{figure}

Example SDTR data on (Zr$_{0.25}$Ta$_{0.25}$Nb$_{0.25}$Ti$_{0.25}$)C is shown in \cref{fig:TCM}. \cref{fig:TCM}(a) shows a full set of spatially-varying thermal wave phase lag data (5, 10, 20, 50, and 100~kHz modulation) for the pristine HEC with an 81~nm Au coating as well as the optimized profile from the forward multi-layer thermal model (dashed line). The retention of the absolute phase delay for the 100~kHz case ensures sensitivity to both the thermal diffusivity and the thermal conductivity of the semi-infinite HEC substrate is retained~\cite{Hurley2015}. \cref{fig:TCM}(b) shows the raw 20~kHz SDTR phase delay data for both the pristine and 25$^\circ$C irradiated specimens. The larger value of the phase delay in the thermal far field for the irradiated sample, at $\sim$10~{\textmu}m, indicates a lower overall thermal diffusivity averaged over the thermal wave penetration depth. Extracting the precise value for the defect-bearing layer thermal conductivity from such a profile requires the use of the multi-layer thermal model described above. 

\begin{table}
\centering
\begin{tabular}{llcc}\hline
 & & Pristine HEC (2-layer) & Defect-bearing HEC (3-layer)\\ \hline
\multicolumn{4}{l}{Layer 1 -- Au transducer} \\ \hline
$R_s$ & [{\textmu}m] & Opt. & Opt. \\ 
$D_1$ & [mm$^2$/s] & 54.7 & 32.1 \\
$\kappa_1$ & [W/m$\cdot$K] & 135.8 & 79.6 \\
$h_1$ & [m] & $81\times10^{-9}$ & $14\times10^{-9}$ \\
$R^{th}_1$ & [m$^2$K/W] & Opt. & Opt. \\ \hline
\multicolumn{4}{l}{Layer 2 -- Pristine bulk or irradiated layer} \\ \hline
$D_2$ & [mm$^2$/s] & Opt. & Opt.$^\dag$ \\
$\kappa_2$ & [W/m$\cdot$K] & Opt. & Opt.$^\dag$ \\
$h_2$ & [m] & $1\times10^{-3}$ & $1.1\times10^{-6}$\\
$R^{th}_2$ & [m$^2$K/W] & -- & 0 \\ \hline
\multicolumn{4}{l}{Layer 3 -- Pristine bulk} \\ \hline
$D_3$ & [mm$^2$/s] & -- & 3.70 \\
$\kappa_3$ & [W/m$\cdot$K] & -- & 9.14 \\
$h_3$ & [m] & -- & $1\times10^{-3}$ \\ \hline
\end{tabular}
\caption{Multi-layer thermal optimization modes for each of the 2- and 3-layer cases considered here. All of the individual parameters optimized are labeled by ``Opt." for that case, while the provided numerical values are taken as fixed parameters. Thermal parameters $D_2$ and $k_2$ are labeled by $^\dag$, for the 3-layer model, as they are optimized self-consistently, ie. their ratio $k_2/D_2=\rho C_p$ is fixed to the value independently optimized for the pristine HEC. In the pristine case, $D_2$ and $k_2$ are optimized independently, allowing $\rho C_p$ to be determined directly from SDTR measurements. The thermal properties of the 81~nm Au coating layer are measured independently using SDTR on a BK7 glass standard, while values for the 14~nm film are inferred using a constitutive model for the conductivity of thin Au films optimized to films generated from the sputter coater used for sample preparation, as these films are too transparent to the heating laser for reliable SDTR measurement~\cite{Dennett2020,Chen1999}.\label{tab:layers}}
\end{table}

\section*{Heat Capacity Measurement using Differential Scanning Calorimetry}

The reference measurement of heat capacity of pristine (Zr$_{0.25}$Ta$_{0.25}$Nb$_{0.25}$Ti$_{0.25}$)C was conducted using a differential scanning calorimeter (DSC 204 F1 Phoenix) with a heating rate of 10~K/min over a temperature range of 40 to 400$^\circ$C in an N$_2$ atmosphere. The reference standard used was Al$_2$O$_3$. As reported in the main manuscript, the representative value at $\sim$335~K for the low end of this temperature range is $315\pm5$~J/kg$\cdot$K.

\section*{Measuring Resistivity and Estimating $\kappa_e$}

\begin{figure}
\centering
\includegraphics[width=0.55\textwidth]{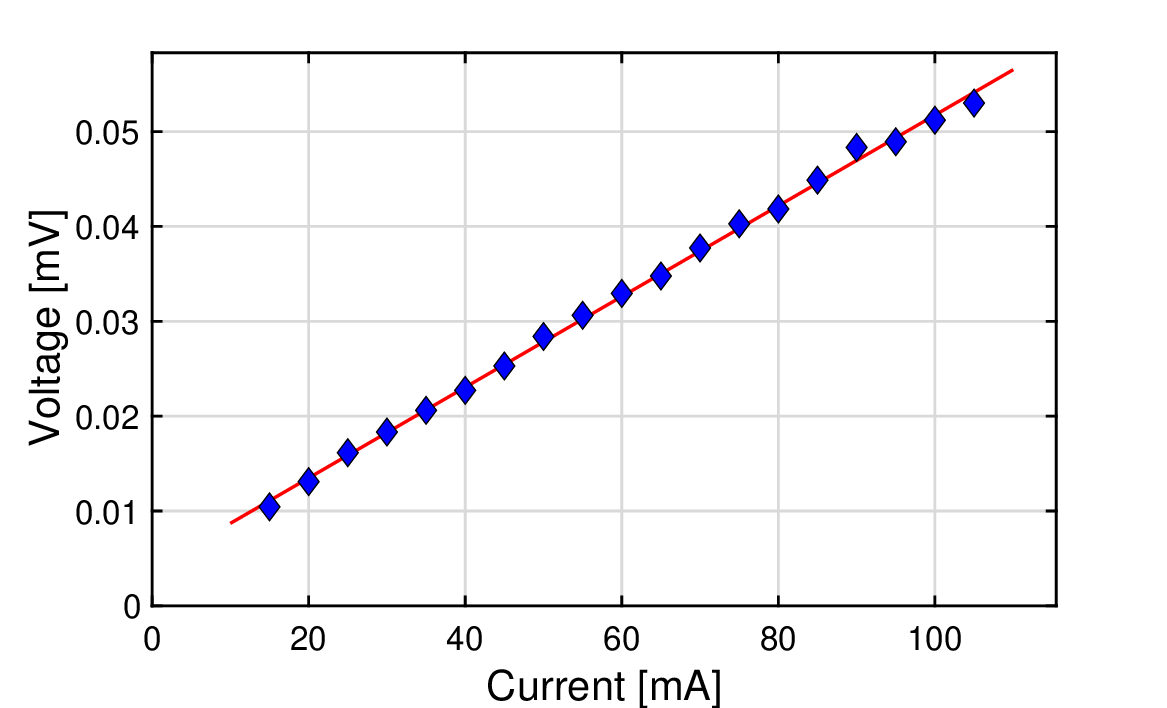}
\caption{Measured (blue symbols) four point probe response on the pristine HEC with best-fit line (red) used to calculate resistivity. The instrument was calibrated using a low-resistivity p-type Si standard in this measurement range prior to dual-polarity measurements at each current indicated.\label{fig:4pt}}
\end{figure}

To estimate the electronic contribution to thermal conductivity for as-synthesized (Zr$_{0.25}$Ta$_{0.25}$Nb$_{0.25}$Ti$_{0.25}$)C, four-point probe measurements were made using a Signatone~Pro4 system using a probe head consisting of four spring-loaded tungsten carbide needles spaced at 0.040'' and a Keithley~2400 source meter. Prior to measurement, the system was calibrated with a p-type, boron doped, silicon calibration standard with electrical resistivity $\rho_e=0.002$~$\Omega$$\cdot$cm from VSLI Standards Incorporated. Per the standard specifications, the system is calibrated at 100~mA, and therefore all resistivity measurements are made in the rage 11-105~mA to remain within a fixed offset range of the Keithley source meter. The specimen measured has a square cross section with side lengths of 4.75~mm and a thickness of 0.99~mm. The voltage drop on the measurement head was recorded in steps of 5~mA by manually controlling the current source in dual-polarity mode and is shown in \cref{fig:4pt}. As-measured voltage drops are converted to resistivity using the standard cross section and thickness corrections as described by Smits~\cite{Smits1958} from a value of V/I extracted from the linear fit shown in \cref{fig:4pt}. In this manner, the electrical resistivity of the as-synthesized 4-cation HEC is measured as $\rho_e=149\pm2$~{\textmu}$\Omega$$\cdot$cm.

As stated in the main manuscript, this measured resistivity is used to estimate the electronic contribution to thermal conductivity as $\kappa_e=LT/\rho_e$. However, as detailed by Rost and coworkers, the Lorentz constant in this family of materials is routinely overestimated by $L=L_0=2.44\times10^{-8}$~W$\Omega$K$^{-2}$~\cite{Rost2020}. As such, we adopt the same procedure described by Rost et al. and compute an effective Lorentz constant, $L_{th}$, as
\begin{equation}
L_{th}=\frac{\frac{\rho_0}{4A}+\left(\frac{T}{\Theta_D}\right)^5J_5\left[\frac{\Theta_D}{T}\right]}{\frac{\rho_0}{4A}+\left(\frac{T}{\Theta_D}\right)^5J_5\left[\frac{\Theta_D}{T}\right]\left(1+\frac{3}{\pi^2}\left(\frac{k_f}{q_D}\right)\left(\frac{\Theta_D}{T}\right)^2-\frac{1}{2\pi^2}\frac{J_7\left[\frac{\Theta_D}{T}\right]}{J_5\left[\frac{\Theta_D}{T}\right]}\right)},
\end{equation}
where $\rho_0$ is a reference ``pristine'' resistivity taken from TaC, $A$ is the impurity contribution for our disordered material evaluated from the measured HEC resistivity $\rho_e$ (from $\rho_e=AT/\Theta_D$), $T$ is the measurement temperature, $\Theta_D$ is the Debye temperature, $k_f$ and $q_D$ are the Fremi and Debye wave vectors, respectively, and $J_n[\Theta_D/T]$ are the Debye integrals defined as
\begin{equation}
J_n\left[\frac{\Theta_D}{T}\right]=\int_0^{\Theta_D/T} dx \frac{x^n\exp[x]}{\left(\exp[x]-1\right)^2}.
\end{equation}
In the calculation of $L_{th}$ a constant ratio $k_f/q_D=2^{-1/3}$ is assumed, taken from free electron theory assuming monovalency and consistent with previous work in refractory carbides and nitrides~\cite{Rost2020,Zheng2017}. We estimate $\Theta_D\simeq739$~K for (Zr$_{0.25}$Ta$_{0.25}$Nb$_{0.25}$Ti$_{0.25}$)C using a Vegard's law average of the Debye temperatures of the constituent binary carbides as listed in \cref{tab:debye}, again following the methods of Rost and coworkers~\cite{Rost2020}. For the measured $\rho_e$, Eq.~(1) returns an ultimate value of $L_{th}=0.76L_0$, which is used to estimate the electronic thermal conductivity as $\kappa_e\simeq3.7$~W/m$\cdot$K for as-synthesized (Zr$_{0.25}$Ta$_{0.25}$Nb$_{0.25}$Ti$_{0.25}$)C.

\begin{table}[h]
\centering
\begin{tabular}{cc}\hline
 & \begin{tabular}{c}Debye \\Temperature [K] \end{tabular}\\ \hline
~~~~NbC~~~~ & 739\\
~~~~TaC~~~~ & 572\\ 
~~~~TiC~~~~ & 942\\
~~~~ZrC~~~~ & 702\\\hline
\end{tabular}
\caption{List of binary constituent carbide Deybe temperatures used for the calculation of $L_{th}$. Values are given as the average of non-\emph{ab~initio} values reported in~\cite{Srivastava2012}.\label{tab:debye}}
\end{table}

\section*{XPS Estimation of Carbon Stoichiometry}

\begin{figure}
\centering
\includegraphics[width=0.47\textwidth]{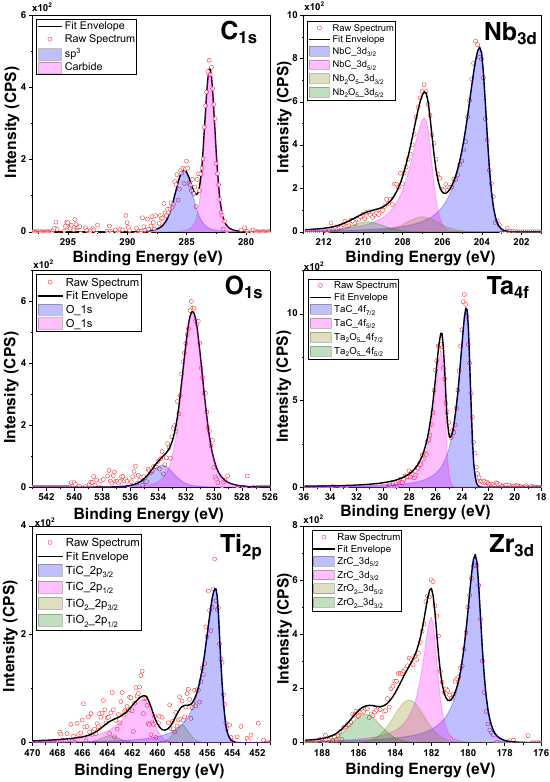}
\caption{Fitted XPS spectra of as-synthesized (Zr$_{0.25}$Ta$_{0.25}$Nb$_{0.25}$Ti$_{0.25}$)C following 10 minutes or Ar ion sputtering. Local energy scans are conducted in an energy band around each of the peaks identified from an initial wide-energy scan (0--1100~eV) maximum sensitivity. \label{fig:xps}}
\end{figure}

X-ray photoemission spectroscopy (XPS) measurements of the sample were collected with a Kratos AXIS Supra XPS system using a monochromatic Al Kalpha source. XPS spectra were collected after 0, 10, and 40 minutes of Ar sputtering on two areas of the as-synthesized specimen: one that had not been coated with an Au transducer and one that was coated for measurement and later had the Au coating mechanically removed. Ar sputtering was performed with an Ar gas cluster source, operated at 5 kV at a 45-degree angle to the specimen. The scans presented in \cref{fig:xps} are taken after 10~minutes of Ar sputtering. The sputtering process creates a 3mm crater in the surface and XPS spectra are collected from a $\sim$110~{\textmu}m diameter spot in the crater. XPS spectra were fit using CASA XPS with a Shirley background subtraction.

For Ar sputtering times of less than 10 minutes, a significant fraction of the metal content is found in a variety of oxide forms due to surface oxidation of the specimen as it was stored in an ambient environments. For 10 minutes of sputtering or greater, the majority of the oxide has been removed with the large majority of the metal appearing bonded to carbon. The metal to carbon ratio of the sample bulk is estimated by taking the ratio of the integrated peak areas of all metals to carbon from four sets of spectra: those collected following 10 and 40 minutes of sputtering in both the uncoated and previously-Au-coated regions. The M:C ratios for each scan are listed in \cref{tab:xps}, and the average metal to carbon ratio is 1.008. However, only a single scan returned M:C$>$1, indicating some spatial heterogeneity remains in the as-synthesized samples.

\begin{table}
\centering
\begin{tabular}{ccc}\hline
Region & Sputtering Time & ~~~M:C~~~ \\ \hline
Uncoated & 10~min & 0.885\\
Uncoated & 40~min & 1.237\\ 
Previously-coated & 10~min & 0.915\\
Previously-coated & 40~min & 0.997\\
Average & -- & 1.008\\\hline
\end{tabular}
\caption{Metal to carbon ratios from each of the four XPS runs conducted on as-synthesized (Zr$_{0.25}$Ta$_{0.25}$Nb$_{0.25}$Ti$_{0.25}$)C.\label{tab:xps}}
\end{table}

\newpage
\bibliography{ref}